\documentclass[12pt]{article}
 \usepackage{epsfig}
 \def\be{\begin{equation}}
 \def\ee{\end{equation}}
 \def\bea{\begin{eqnarray}}
 \def\eea{\end{eqnarray}}
 \usepackage{graphicx}

 \catcode`\@=11
 \def\lsim{\mathrel{\mathpalette\@versim<}}
 \def\gsim{\mathrel{\mathpalette\@versim>}}
 \def\@versim#1#2{\vcenter{\offinterlineskip
 \ialign{$\m@th#1\hfil##\hfil$\crcr#2\crcr\sim\crcr } }}
 \catcode`\@=12

 \catcode`@=12
\begin{document}
 \thispagestyle{empty}
 \begin{flushright}
 UCRHEP-T593\\
 Aug 2018\
 \end{flushright}
 \vspace{0.6in}
 \begin{center}
 {\LARGE \bf Baryon-Lepton Duplicity as the\\ Progenitor of Long-Lived 
Dark Matter\\}
 \vspace{1.2in}
 {\bf Ernest Ma\\}
 \vspace{0.2in}

{\sl Physics and Astronomy Department,\\ 
University of California, Riverside, California 92521, USA\\}
\vspace{0.1in}
{\sl Jockey Club Institute for Advanced Study,\\ 
Hong Kong University of Science and Technology, Hong Kong, China\\} 
\end{center}
 \vspace{1.2in}

\begin{abstract}\
In an $SU(2)_R$ extension of the standard model, it is shown how the 
neutral fermion $N$ in the doublet $(N,e)_R$ may be assigned baryon 
number $B=1$, in contrast to its $SU(2)_L$ counterpart $\nu$ in the 
doublet $(\nu,e)_L$ which has lepton number $L=1$.  This baryon-lepton 
duplicity allows a scalar $\sigma$ which couples to ${N}_L N_L$ to be 
long-lived dark matter.
\end{abstract}

 \newpage
 \baselineskip 24pt

\noindent \underline{\it Introduction}~:~
In the conventional $SU(2)_R$ extension of the standard model (SM) of 
quarks and leptons, the neutral fermion $N$ in the doublet $(N,e)_R$ is 
identified with the Dirac mass partner of $\nu$ in the $SU(2)_L$ 
doublet $(\nu,e)_L$.  Hence $N$ has lepton number $L=1$.  However, it 
is also possible that $N$ is not the mass partner of $\nu$, and that 
it has $L=0$~\cite{klm09} or $L=2$~\cite{klm10}.  As such $N$ may be 
considered a dark-matter candidate using~\cite{m15} $(-1)^{L+2j}$ as the 
stabilizing dark symmetry.  In the following, it will be shown how 
$N$ may be assigned baryon number $B=1$ instead~\cite{m88}, in which 
case a scalar $\sigma$ coupling to ${N}_L N_L$ may become long-lived 
dark matter.

\noindent \underline{\it Model}~:~
The basic framework for considering $(N,e)_R$ differently from $(\nu,e)_R$ 
is originally inspired by $E_6$ models with the decomposition 
$E_6 \to SU(3)_C \times SU(3)_L \times SU(3)_R \to SU(3)_C \times SU(2)_L 
\times SU(2)_R \times U(1)_X$, where the $SU(2)_R$ is not~\cite{m87} the 
one contained in $SO(10) \to SU(5)$ in the conventional left-right model. 
Consider the fermion particle content of the basic model given in Table 1.
\begin{table}[htb]
\caption{Basic fermion content of model with unconventional $SU(2)_R$.}
\begin{center}
\begin{tabular}{|c|c|c|c|c|c|}
\hline
fermion & $SU(3)_C$ & $SU(2)_L$ & $SU(2)_R$ & $U(1)_X$ & $Z_5$ \\
\hline
$(\nu,e)_L$ & 1 & 2 & 1 & $-1/2$ & 1 \\ 
$\nu_R$ & 1 & 1 & 1 & 0 & 1 \\ 
\hline
$(N,e)_R$ & 1 & 1 & 2 & $-1/2$ & $\omega^{-1}$ \\ 
$N_L$ & 1 & 1 & 1 & 0 & $\omega^3$ \\ 
\hline
$(u,d)_L$ & 3 & 2 & 1 & 1/6 & $\omega$ \\ 
$d_R$ & 3 & 1 & 1 & $-1/3$ & $\omega$ \\ 
\hline
$(u,h)_R$ & 3 & 1 & 2 & 1/6 & $\omega^2$ \\ 
$h_L$ & 3 & 1 & 1 & $-1/3$ & $\omega^{-2}$ \\ 
\hline
\end{tabular}
\end{center}
\end{table}

The electric charge is $Q = I_{3L} + I_{3R} + X$.  The discrete $Z_5$ 
symmetry ($\omega^5=1$) serves to forbid the terms $\bar{N}_L \nu_R$ and 
$\bar{h}_L d_R$ and others to be discussed.  The scalar particle content 
of the proposed model of baryon-lepton duplicity is given in Table 2.
\begin{table}[htb]
\caption{Scalar content defining the model of baryon-lepton duplicity.}
\begin{center}
\begin{tabular}{|c|c|c|c|c|c|}
\hline
scalar & $SU(3)_C$ & $SU(2)_L$ & $SU(2)_R$ & $U(1)_X$ & $Z_5$ \\
\hline
$(\phi_L^+,\phi_L^0)$ & 1 & 2 & 1 & $1/2$ & 1 \\ 
$(\phi_R^+,\phi_R^0)$ & 1 & 1 & 2 & $1/2$ & $\omega^{-1}$ \\ 
\hline
$\eta$ & 1 & 2 & 2 & 0 & $\omega$ \\ 
\hline
$\zeta$ & 3 & 1 & 1 & $-1/3$ & $\omega^{-2}$ \\ 
\hline
$\sigma$ & 1 & 1 & 1 & 0 & $\omega$ \\
\hline
\end{tabular}
\end{center}
\end{table}

In the above, $\eta$ is a bidoublet, with $SU(2)_L$ acting vertically and 
$SU(2)_R$ acting horizontally, i.e.
\begin{eqnarray}
\eta &=& \pmatrix{\eta_1^0 & \eta_2^+ \cr \eta_1^- & \eta_2^0}, \\ 
\tilde{\eta} &=& \sigma_2 \eta^* \sigma_2 = \pmatrix{\bar{\eta}_2^0 & 
-\eta_1^+ \cr -\eta_2^- & \bar{\eta}_1^0}.
\end{eqnarray}
The $Z_5$ symmetry distinguishes $\eta$ from $\tilde{\eta}$. 
The resulting Yukawa interactions are
\begin{eqnarray}
{\cal L}_Y &=& f_\nu(\bar{\nu}_L \bar{\phi}^0_L - \bar{e}_L \phi^-_L)\nu_R 
+ f_d(\bar{u}_L \phi^+_L + \bar{d}_L \phi_L^0)d_R \nonumber \\ 
&+& f_N(\bar{N}_R \bar{\phi}_R^0 - \bar{e}_R \phi_R^-)N_L 
+ f_h(\bar{u}_R \phi_R^+ + \bar{h}_R \phi_R^0)h_L \nonumber \\ 
&+& f_e[(\bar{\nu}_L \eta_1^0 + \bar{e}_L \eta_1^-)N_R + (\bar{\nu}_L \eta_2^+ 
+\bar{e}_L \eta_2^0)e_R] \nonumber \\ 
&+& f_u[(\bar{u}_L \bar{\eta}_2^0 - \bar{d}_L \eta_2^-)u_R + (-\bar{u}_L 
\eta_1^+ + \bar{d}_L \bar{\eta}_1^0)h_R] \nonumber \\ 
&+& f_1 \sigma^* N_L N_L + 
f_2 \bar{N}_L d_R \zeta^* + f_3 \bar{\nu}_R h_L \zeta^*\nonumber \\ 
&+& f_4 \epsilon_{ijk}(u_{iL} d_{jL} - d_{iL} u_{jL}) \zeta_k + H.c.,
\end{eqnarray}
where each $f$ is a $3 \times 3$ matrix for the three families of quarks 
and leptons and the last term is the product of three color triplets.
Note that $\nu$ and $d$ masses come from $\langle \phi_L^0 \rangle$, 
$e$ and $u$ masses come from $\langle \eta_2^0 \rangle$, $N$ and 
$h$ masses come from $\langle \phi_R^0 \rangle$.  This structure 
guarantees the absence of tree-level flavor-changing neutral currents.

If the scalar color triplet $\zeta$ and singlet $\sigma$ are absent, the 
model of Ref.~\cite{klm10} 
is recovered with $L=2$ for $N$ and $L=-1$ for $h$.  As it is, a very 
different outcome is obtained with $B=1$ for $N$ and $B=2$ for $\sigma$ as 
well as $B=-2/3$ for $h$, as shown below.  The first thing to realize is 
that even though the input symmetry is $Z_5$, the Lagrangian of Eq.~(3) 
actually has a larger symmetry due to the chosen particle content under 
the gauge symmetry.  It is an $U(1)$ symmetry $S$ under which
\begin{eqnarray}
&& (u,d)_L, d_R \sim 1/3,~~~h_L \sim -2/3,~~~(u,h)_R \sim -1/6, ~~~
N_L \sim 1, ~~~ (N,e)_R \sim 1/2,\\ 
&&  \Phi_R \sim 1/2, ~~~ \eta \sim -1/2, 
~~~ \sigma \sim 2, ~~~ \zeta \sim -2/3.
\end{eqnarray}
This $S$ symmetry is broken by $\langle \phi_R^0 \rangle$ as well as 
$\langle \eta_2^0 \rangle$, but not the combination $S + I_{3R}$. 
Indeed this residual symmetry is just baryon number, i.e. 1/3 for 
the known quarks and zero for the known leptons.  There is another 
residual symmetry, i.e. lepton parity under which the known leptons 
are odd.  Note that $\nu_R$ is allowed a Majorana mass, hence the 
canonical seesaw mechanism for neutrino mass is applicable.

As for the new particles beyond the SM, their baryon number and lepton 
parity assignments are
\begin{eqnarray}
&& \zeta \sim (-2/3,+), ~~~ N \sim (1,+), ~~~ \sigma \sim (2,+), 
~~~ h \sim (-2/3,-), \\
&& W_R^\pm \sim (\pm 1,-), ~~~ Z' \sim (0,+), ~~~ (\eta_1^0,\eta_1^-) 
\sim (-1,-).
\end{eqnarray}
Hence $\zeta$ is a scalar diquark, $h$ is a fermion diquark with odd lepton 
parity, $N$ is a fundamental $B=1$ fermion, 
$\sigma$ is a fudamental $B=2$ scalar, 
$W_R^+$ is a fundamental $B=1$ vector boson with odd lepton parity, 
and $(\eta_1^+, -\bar{\eta}^0_1)$ is a fundamental $B=1$ scalar $SU(2)_L$ 
doublet with odd lepton parity.  Underlying this exotic scenario is the 
duplicity between $N$ and $\nu$ in their $SU(2)_R/SU(2)_L$ interactions. 
Two symmetries are conserved: baryon number and lepton parity.  The 
lightest lepton, i.e. the lightest neutrino, is stable.  The lightest 
baryon, i.e. the proton, is stable.  However, just as the heavier neutrinos 
are very long-lived, the heavier $B=2$ $\sigma$ may also be very long-lived 
and become dark matter.  

\noindent \underline{\it Gauge Boson Masses and Interactions}~:~
Let
\begin{equation}
\langle \phi^0_L \rangle = v_1, ~~~ \langle \eta^0_2 \rangle = v_2, ~~~ 
\langle \phi^0_R \rangle = v_R, 
\end{equation}
then the $SU(3)_C \times SU(2)_L \times SU(2)_R \times U(1)_X$ 
gauge symmetry is broken to $SU(3)_C \times U(1)_Q$, with residual baryon 
number and lepton parity as discussed in the previous section.  
Consider now the masses of the gauge bosons.  The charged ones, $W_L^\pm$ 
and $W_R^\pm$, do not mix because of $B$ and $(-1)^L$, as in the original 
alternative left-right models.  Their masses are given by
\begin{equation}
M_{W_L}^2 = {1 \over 2} g_L^2 (v_1^2 + v_2^2), ~~~ 
M_{W_R}^2 = {1 \over 2} g_R^2 (v_R^2 + v_2^2).
\end{equation}
Since $Q = I_{3L} + I_{3R} + X$, the photon is given by
\begin{equation}
A = {e \over g_L} W_{3L} + {e \over g_R} W_{3R} + {e \over g_X} X,
\end{equation}
where $e^{-2} = g_L^{-2} + g_R^{-2} + g_X^{-2}$.  Let
\begin{eqnarray}
Z &=& (g_L^2 + g_Y^2)^{-1/2} \left( g_L W_{3L} - {g_Y^2 \over g_R} W_{3R} 
- {g_Y^2 \over g_X} X \right), \\ 
Z' &=& (g_R^2 + g_X^2)^{-1/2} ( g_R W_{3R} - g_X X),
\end{eqnarray}
where $g_Y^{-2} = g_R^{-2} + g_X^{-2}$, 
then the $2 \times 2$ mass-squared matrix spanning $(Z,Z')$ has the 
entries:
\begin{eqnarray}
M^2_{ZZ} &=& {1 \over 2} (g_L^2 + g_Y^2) (v_1^2 + v_2^2), \\ 
M^2_{Z'Z'} &=& {1 \over 2} (g_R^2 + g_X^2) v_R^2 + {g_X^4 v_1^2 + g_R^4 v_2^2 
\over 2(g_R^2 + g_X^2)}, \\ 
M^2_{ZZ'} &=& {\sqrt{g_L^2 + g_Y^2} \over 2\sqrt{g_R^2 + g_X^2}} 
(g_X^2 v_1^2 - g_R^2 v_2^2).
\end{eqnarray}
Their neutral-current interactions are given by
\begin{eqnarray}
{\cal L}_{NC} = e A_\mu j^\mu_Q + g_Z Z_\mu (j^\mu_{3L} - \sin^2 \theta_W 
j^\mu_Q) + (g_R^2 + g_X^2)^{-1/2} Z'_\mu (g_R^2 j^\mu_{3R} - g_X^2 j^\mu_X),
\end{eqnarray}
where $g_Z^2 = g_L^2 + g_Y^2$ and $\sin^2 \theta_W = g_Y^2/g_Z^2$.
Assuming also that $g_R = g_L$, then 
$g_X^2 /g_Z^2 = \sin^2 \theta_W \cos^2 \theta_W / \cos 2 \theta_W$. 
In that case, setting $v_1^2/v_2^2 = \cos 2 \theta_W/\sin^2 \theta_W$ 
would result in zero $Z-Z'$ mixing which is constrained by precision 
data to be less than a few times $10^{-4}$.

The present bound on $M_{Z'}$ from the Large Hadron Collider (LHC) is 
about 4 TeV.  However, if the lightest $N$ is considered as dark matter, 
then its gauge interaction through $Z'$ with quarks would constrain 
$M_{Z'}$ to be above 10 TeV or higher from direct-search 
experiments, depending on $m_N$.  Here it will be assumed that $\sigma$ 
is dark matter 
and since it does not couple to $Z'$, this constraint is not applicable.

\noindent \underline{\it Scalar Sector}~:~
Consider the most general scalar potential consisting of $\Phi_{L,R}$ 
and  $\eta$, i.e.
\begin{eqnarray}
V &=& -\mu^2_L \Phi_L^\dagger \Phi_L - \mu^2_R \Phi_R^\dagger \Phi_R - 
\mu^2_\eta Tr(\eta^\dagger \eta) 
+ [\mu_3 \Phi_L^\dagger \eta \Phi_R + H.c.] 
+ {1 \over 2} \lambda_L (\Phi_L^\dagger \Phi_L)^2 \nonumber \\ 
&+& {1 \over 2} \lambda_R 
(\Phi_R^\dagger \Phi_R)^2 +  
{1 \over 2} \lambda_\eta [Tr(\eta^\dagger \eta)]^2 + {1 \over 2} \lambda'_\eta 
Tr(\eta^\dagger \eta \eta^\dagger \eta) 
+ \lambda_{LR} (\Phi_L^\dagger \Phi_L)(\Phi_R^\dagger \Phi_R) \nonumber \\ 
&+& \lambda_{L\eta} \Phi_L^\dagger \eta \eta^\dagger \Phi_L + \lambda'_{L\eta} 
\Phi_L^\dagger \tilde{\eta} \tilde{\eta}^\dagger \Phi_L + \lambda_{R\eta} 
\Phi_R^\dagger \eta^\dagger \eta \Phi_R + \lambda'_{R\eta} 
\Phi_R^\dagger \tilde{\eta}^\dagger \tilde{\eta} \Phi_R.
\end{eqnarray}
Note that
\begin{eqnarray}
2 |det(\eta)|^2 &=& [Tr(\eta^\dagger \eta)]^2 - Tr(\eta^\dagger \eta 
\eta^\dagger \eta), \\ 
(\Phi_L^\dagger \Phi_L) Tr(\eta^\dagger \eta) &=& \Phi_L^\dagger \eta 
\eta^\dagger 
\Phi_L + \Phi_L^\dagger \tilde{\eta} \tilde{\eta}^\dagger \Phi_L, \\ 
(\Phi_R^\dagger \Phi_R) Tr(\eta^\dagger \eta) &=& \Phi_R^\dagger 
\eta^\dagger \eta 
\Phi_R + \Phi_R^\dagger \tilde{\eta}^\dagger \tilde{\eta} \Phi_R.
\end{eqnarray} 
The minimum of $V$ satisfies the conditions
\begin{eqnarray}
\mu_L^2 &=& \lambda_L v_1^2 + \lambda_{L\eta} v_2^2 + \lambda_{LR} v_R^2  
+ \mu_3 v_2 v_R/v_1, \\ 
\mu_\eta^2 &=& (\lambda_\eta + \lambda'_\eta) v_2^2 + \lambda_{L\eta} v_1^2 + 
\lambda_{R\eta} v_R^2 + \mu_3 v_1 v_R/v_2, \\ 
\mu_R^2 &=& \lambda_R v_R^2 + \lambda_{LR} v_1^2 + \lambda_{R\eta} v_2^2  
+ \mu_3 v_1 v_2/v_R.
\end{eqnarray}
The $3 \times 3$ mass-squared matrix spanning $\sqrt{2}Im(\phi_L^0,\eta_2^0,
\phi_R^0)$ is then given by
\begin{eqnarray}
{\cal M}^2_I = \mu_3 \pmatrix{-v_2v_R/v_1 & v_R & v_2  \cr v_R & 
-v_1v_R/v_2 & -v_1 \cr v_2 & -v_1 & -v_1v_2/v_R}.
\end{eqnarray}
and that spanning $\sqrt{2}Re(\phi_L^0,\eta_2^0,\phi_R^0)$ is 
${\cal M}^2_R =$ 
\begin{eqnarray}
\mu_3 \pmatrix{-v_2v_R/v_1 & v_R & v_2 \cr v_R & 
-v_1v_R/v_2 & v_1 \cr v_2 & v_1 & -v_1v_2/v_R} 
+ 2\pmatrix{ \lambda_L v_1^2 & \lambda_{L\eta}v_1v_2 &  
\lambda_{LR}v_1 v_R  \cr \lambda_{L\eta}v_1v_2 
& (\lambda_\eta + \lambda'_\eta)v_2^2 & \lambda_{R\eta}v_2v_R 
 \cr  \lambda_{LR} v_1 v_R &  \lambda_{R\eta} 
v_2 v_R &  \lambda_R v_R^2}.
\end{eqnarray}
Hence there are two zero eigenvalues in ${\cal M}^2_I$ with one nonzero 
eigenvalue $-\mu_3[v_1 v_2/v_R + v_R (v_1^2+v_2^2)/v_1v_2]$ corresponding 
to the eigenstate 
$A = (-v_1^{-1},v_2^{-1},v_R^{-1})/\sqrt{v_1^{-2}+v_2^{-2}+v_R^{-2}}$. 
In ${\cal M}^2_R$, the linear combination 
$H = (v_1,v_2,0)/\sqrt{v_1^2+v_2^2}$, 
is the standard-model Higgs boson, with 
\begin{equation}
m_H^2 = 2[\lambda_L v_1^4 + (\lambda_\eta + \lambda'_\eta) v_2^4 
+ 2 \lambda_{L\eta} v_1^2 v_2^2]/(v_1^2 + v_2^2).
\end{equation}
The other two scalar bosons, i.e. $H' = (v_2,-v_1,0)\sqrt{v_1^2+v_2^2}$ and 
$H_R = (0,0,1)$ are assumed not to mix with $H$ and each other by 
fine-tuning the $\lambda$ parameters to avoid further experimental 
constraints. 

There are four charged scalars in the Higgs potential of Eq.~(17).  Two 
have $B=1$, i.e. $\phi_R^+,\eta_1^+$.  One linear combination becomes 
the longitudinal component of $W_R^+$.  The orthogonal combination, i.e. 
$(v_R \eta_1^+ - v_2 \phi_R^+)/\sqrt{v_R^2+v_2^2}$ has a mass given by
\begin{equation}
m^2 = (\lambda'_{R\eta} - \lambda_{R\eta})(v_R^2 + v_2^2) - {\mu_3 v_1 
(v_R^2 + v_2^2) \over v_R v_2}.
\end{equation}
The other two charged scalars have $B=0$, i.e. $\phi_L^+,\eta_2^+$. 
One linear combination becomes the longitudinal component of $W_L^+$.  
The orthogonal combination, i.e. 
$(v_1 \eta_2^+ - v_2 \phi_L^+)/\sqrt{v_1^2+v_2^2}$ has a mass given by
\begin{equation}
m^2 = (\lambda'_{L\eta} - \lambda_{L\eta})(v_1^2 + v_2^2) - {\mu_3 v_R 
(v_1^2 + v_2^2) \over v_1 v_2}.
\end{equation}
The two physical charged scalars do not mix, for the same reason that 
$W_R$ and $W_L$ do not mix, because they have different $B$ values.
Note that in the limit $v_{1,2} << v_R$, the $B=0$ charged scalar has 
the same mass as the $B=0$ scalar $H'$ and the $B=0$ pseudoscalar $A$, 
as expected.  As for the $B=1$ neutral scalar $\eta_1^0$, it has a mass 
given by 
\begin{eqnarray}
m^2(\eta_1^0) = (\lambda'_{R\eta}-\lambda_{R\eta})v_R^2 + 
(\lambda'_{L\eta}-\lambda_{L\eta})v_1^2 - \lambda'_\eta v_2^2 
- \mu_3(v_1/v_2)v_R.
\end{eqnarray} 
Note again that in the limit $v_{1,2} << v_R$, The $B=1$ charged and 
neutral scalars have the same mass, as expected.

\noindent \underline{\it Diquark Connection to Dark Matter}~:~
The scalar diquark $\zeta$ is crucial in assigning $B=1$ to $N$ and 
$\sigma$.  The decay of $N_L$ to $d_R$ and a virtual $\zeta^*$ which 
converts to $u_L d_L$ means that $N$ is long-lived if $m_\zeta$ is very 
large.  The current LHC bound on $m_\zeta$ is about 2.5 TeV.  
On the other hand, if $m_\sigma < m_N$, then the former's decay is 
even more suppressed.  It will be shown how $\sigma$ may indeed be 
suitable as long-lived dark matter.

Of the new particles with $B \neq 0$, $\zeta$ is assumed to be the heaviest 
and $N$ to be the lightest except $\sigma$. 
Now $N$ decays to $udd$ with a decay rate given by~\cite{abgh02}
\begin{equation}
\Gamma(N \to udd) = {f_2^2 f_4^2 m_N^5 \over 8 (4\pi)^3 m_\zeta^4} 
\int_0^1 d \lambda_2 \lambda_2(1-\lambda_2)^2 = 
{f_2^2 f_4^2 m_N^5 \over 96 (4\pi)^3 m_\zeta^4}.
\end{equation}
As an example, a lifetime of 
\begin{equation}
\tau_N = (5 \times 10^{24}~s) \left( {0.01 \over f_2} \right)^2 
\left( {0.01 \over f_4} \right)^2 \left( {300~{\rm GeV} \over m_N} 
\right)^5 \left( {m_\zeta \over 10^{12}~{\rm GeV}} \right)^4
\end{equation}
is obtained.  Note that the age of the Universe is $4.35 \times 10^{17}s$, 
but the bound on decaying dark matter~\cite{sw17} is much greater, say 
about $10^{25}s$, 
from the constraint of the cosmic microwave background (CMB).  Hence $N$ 
may be long-lived enough for it to be dark matter, with some adjustment 
of parameter values.  However, because its 
interactions through the new gauge boson $Z'$ are constrained by 
direct-search data as pointed out already, its resulting annihilation 
cross section is too small and would result in a thermal relic abundance 
exceeding what is observed.  Hence it will be assumed from now on that 
$N$ decays quickly, using for example $m_\zeta = 10^5$ GeV so that 
$\tau_N = 5 \times 10^{-4}s$ which is certainly short enough not to 
disturb Big Bang Nucleosynthesis (BBN).

Excepting $\sigma$, the other new particles with $B \neq 0$ all decay 
quickly to $N$.  The vector gauge boson $W_R^-$ decays to $e^- \bar{N}$. 
The fermion diquark $h$ decays to $u W_R^-$ if $m_h > M_{W_R}$, or to 
$u e^- \bar{N}$ if $m_h < M_{W_R}$.  The $B=-1$ scalar doublet 
$(\eta_1^0,\eta_1^-)$ decays to the $B=0$ scalar doublet 
$(\eta_2^+,\eta_2^0)$ through $W_R^-$, again converting to $e^- \bar{N}$.  
In all these decays, $N$ would appear as missing energy because its 
lifetime is long enough to escape detection in the experimental apparatus.

To estimate the decay rate of $\sigma \to NN$, let $p_{1,2}$ be the 
sum of the four-momenta of the three quark jets from each $N$.  Then 
\begin{equation}
\Gamma_\sigma \sim {m_\sigma \over 16 \pi} \left[ {f_1 f_2^2 f_4^2 \over 
96 (4\pi)^3 m_\zeta^4} \right]^2 \int dp_1^2 dp_2^2 {(m_\sigma -p_1^2-p_2^2)^2 
p_1^4 p_2^4 \over (p_1^2-m_N^2)^2 (p_2^2-m_N^2)^2},
\end{equation}
where $p^2_{1,2} > 0$ and $p_1^2+p_2^2 < m_\sigma^2$.  Letting 
$p_1^2=p_2^2=m_\sigma^2$ in the denominator, the integral is bounded 
from above by
\begin{eqnarray}
{1 \over (m_N^2-m_\sigma^2)^4} \int_0^{m_\sigma^2} p_1^4 dp_1^2 
\int_0^{m_\sigma^2-p_1^2} p_2^4 (m_\sigma^2-p_1^2-p_2^2)^2 dp_2^2 
=  {m_\sigma^{16} \over 5040(m_N^2-m_\sigma^2)^4}.
\end{eqnarray}
Hence
\begin{equation}
\Gamma_\sigma < {m_\sigma^{17} \over 35 \pi (m_N^2-m_\sigma^2)^4} 
\left[ {f_1 f_2^2 f_4^2 \over 9(32\pi)^3 m_\zeta^4} \right]^2.
\end{equation}
For $m_\zeta = 10^5$ GeV, $f_{1,2,4}=0.01$, $m_N = 300$ GeV, and 
$m_\sigma = 250$ GeV, $\tau_\sigma > 6 \times 10^{28}s$ is obtained. 
This shows that $\sigma$ is long-lived enough to be dark matter.

\noindent \underline{\it Relic Abundance of $\sigma$}~:~
\begin{figure}[htb]
\vspace*{0.5cm}
\hspace*{1cm}
\includegraphics[scale=0.8]{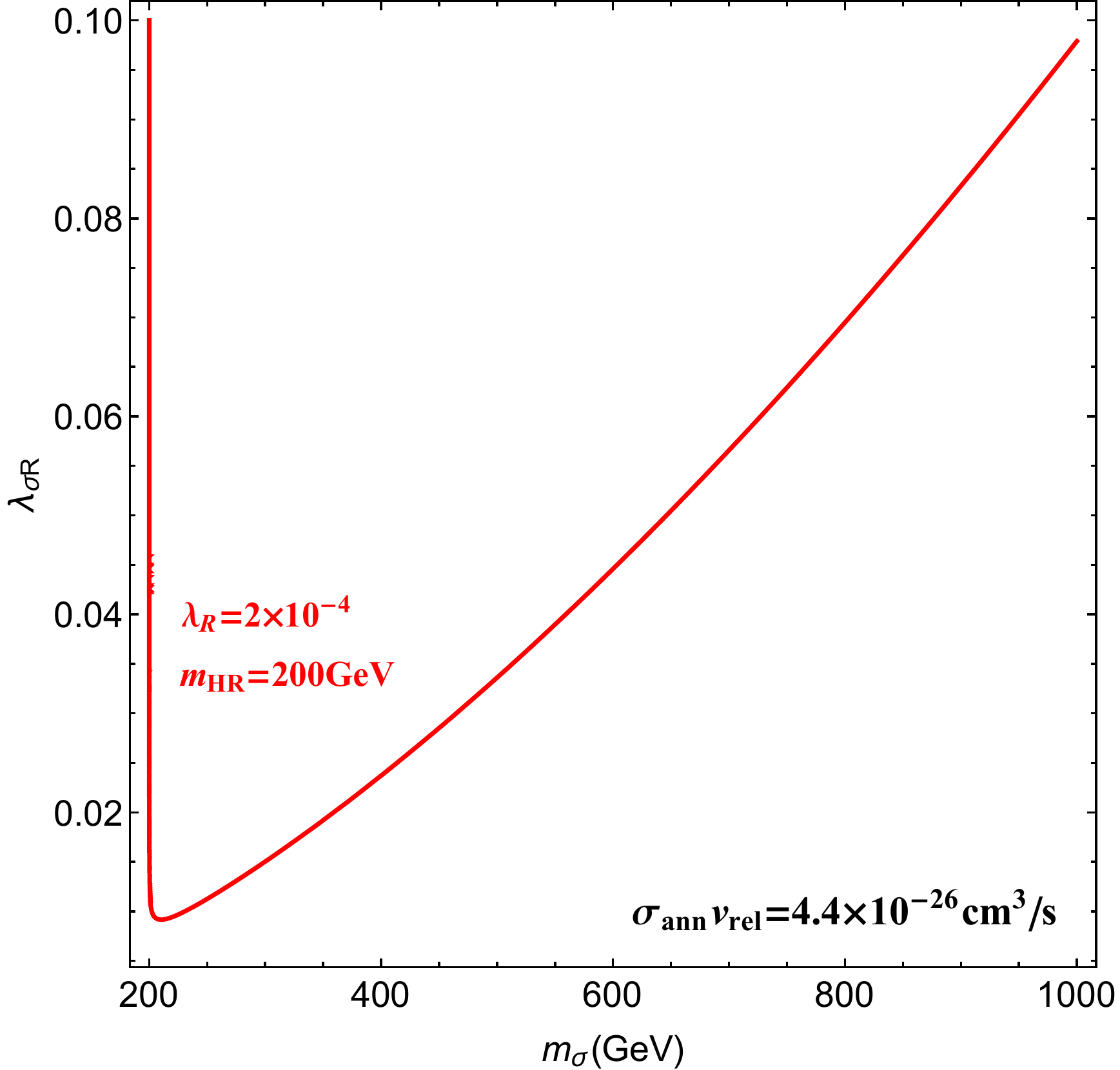}
\caption{$\lambda_{\sigma R}$ verus $m_\sigma$ for $\sigma$ relic abundance.}
\end{figure}
The quartic interaction coupling $\lambda_{\sigma H}$ with the SM 
Higgs boson must be small ($< 10^{-3}$) to avoid the constraint of 
direct-search experiments.  This means that the $\sigma \sigma^* \to HH$ 
cross section is not large enough to obtain the correct thermal relic 
abundance for $\sigma$ as dark matter.  However, if the $H_R$ boson 
is lighter than $\sigma$, the latter's annihilation to the former (which 
does not couple to SM fermions) is a possible mechanism.  Using
\begin{equation}
\sigma_{ann} \times v_{rel} = {\lambda_{\sigma R}^2 r \sqrt{1-r} \over 
64 \pi m^2_{H_R}} \left[ 1 + {3r \over 4-r} - \left( {\lambda_{\sigma R} \over 
\lambda_R} \right) {r \over 2-r} \right]^2,
\end{equation}
where $r = m^2_{H_R}/m^2_\sigma$ and assuming as an example 
$\lambda_R = 2 \times 10^{-4}$ 
with $v_R = 10$ TeV, so that $m_{H_R} = 200$ GeV, the allowed range of 
$\lambda_{\sigma R}$ is plotted versus $m_\sigma = 200~{\rm GeV}/\sqrt{r}$ 
in Fig.~1 for $\sigma_{ann} \times v_{rel} = 4.4 \times 10^{-26}~cm^2/s$. 

\noindent \underline{\it Concluding Remarks}~:~
In an $SU(2)_R$ extension of the SM, where an input $Z_5$ discrete symmetry 
is imposed with the particle content of Table 1 and Table 2, it has been shown 
that two conserved symmetries emerge. One is lepton parity $(-1)^L$ so that 
the known leptons are odd and other SM particles are even.  Neutrino masses 
are obtained through the usual canonical seesaw mechanism.  The other is 
baryon number $B$ with the usual assignment of 1/3 for the SM quarks. 
The conservation of $B$ and $(-1)^L$ separately implies that the proton is 
stable.

What is new and unconventional in this model is the nature of the neutral 
fermion $N$ in the $SU(2)_R$ doublet $(N,e)_R$.  It is not the Dirac mass 
partner of the neutrino $\nu$ in the $SU(2)_L$ doublet $(\nu,e)_L$.  Instead 
of having $L=1$, it actually has $B=1$, as explained in the text 
because of the $Z_5$ symmetry and the chosen particle content. 
This baryon-lepton duplicity allows new particles to have nonzero $B$ 
as well as odd $(-1)^L$.  Whereas $N$ itself decays into three quark jets 
and may have a long lifetime, a scalar $\sigma$ with $B=2$ and 
$m_\sigma < m_N$ is proposed instead as dark matter with 
a lifetime many orders of magnitude exceeding the age of the Universe. 
It has a correct thermal relic abundance from its interaction with the Higgs 
boson $H_R$ associated with $SU(2)_R$ symmetry breaking.  Its interaction 
with the $SU(2)_L$ Higgs boson $H$ is however adjustable, so that 
present direct-search bounds are obeyed, but may reveal itself in the 
future if a positive signal is measured.

To test this model, the $SU(2)_R$ gauge sector has to be probed.  If $W_R$ 
or $Z'$ can be produced, then $N$ is predicted as a decay product. 
It will appear as an invisible massive particle.  Another prediction is 
the existence of the fermion diquark $h$ with odd lepton parity.  It 
is produced readily by gluon interactions at the LHC and decays to 
$u e^- \bar{N}$ which looks like a fourth-family quark but again 
$N$ appears as an invisible massive particle and not the expected 
light neutrino.  As the LHC gathers more data, these processes 
may be searched for.

\noindent \underline{\it Acknowledgement}~:~
This work was supported in part by the U.~S.~Department of Energy Grant 
No. DE-SC0008541.

\bibliographystyle{unsrt}

\end{document}